# Refractive index sensing with hybrid surfaces of photonic crystals and dielectric microsphere monolayers

Cheng Fang[1], Qingqing Shang[1], Fen Tang[2], Songlin Yang[3], and Ran Ye[2,*]

[1]School of Physics and Technology, Nanjing Normal University, Nanjing 21033, China
[2]School of Computer and Electronic Information, Nanjing Normal University, Nanjing 21033, China
[3]Advanced Photonics Center, Southeast University, Nanjing 21033, China

*Corresponding: ran.ye@njnu.edu.cn*

## Abstract

In this work, a refractive index (RI) sensor with an effective integration of colorimetric detection and optical sensing capabilities has been developed. Colorimetric detection relies on the sensitivity of the structural color of photonic crystal (PC) substrates to the changes in background RI, while the optical sensing is performed by measuring the magnification abilities of the dielectric microspheres, which depends on the position of the photonic nanojet. Based on this concept, we have successfully assembled 35 $\mu$m-diameter barium titanate glass microspheres, 4.9 $\mu$m-diameter polystyrene and silica microsphere monolayers on 1D or 2D PC substrates to perform RI sensing in various liquids. In addition, the developed RI sensor is highly compatible with commercial optical microscopes and applicable for RI sensing in areas as small as tens of square microns.

## 1   Introduction

Photonic nanojet (PNJ) effect is the extraordinary focusing of electromagnetic waves in small dielectric particles with wavelength-scale dimensions [1, 2]. It can be generated with transparent materials like organic molecules [3, 4], silica [5–7], barium titanate glass (BTG) [8], etc. The characteristics of PNJ depend both on the properties of dielectric particles and the surrounding environment. For example, a PNJ will become longer if the illumination source gets closer to a particle [9] or if the particle diameter is increased [10]. As for the sharpness of a PNJ, it was reported that a dielectric microsphere with a higher RI will generate a PNJ with a smaller Full-Width Half-Maximum (FWHM) value [11]. In addition, coating BTG microspheres with polystyrene (PS) thin films will also lead to a narrower PNJ [12]. Dielectric particles can generate a PNJ inside the particle, close to the particle surface or far away from the particle. The distance between a PNJ and the particle is very sensitive to the surrounding environment. It was found that the PNJ position, or the focal length (FL) of a dielectric particle, is highly dependent on the RI contrast between the particle itself and the background medium [11, 13]. With a constant particle RI, any fluctuations in the surrounding environment will immediately affect the PNJ position, and therefore affect the imaging performance of the microspheres [14]. In recent years, the development of PNJ-based technologies has attracted great attention, leading to the emergence of microsphere-assisted super-resolution microscopy [15, 16], high-resolution laser micromachining [17], surface plasmon propagation enhancement [18], visualized nanomanipulation based on scanning microsphere systems [19], etc.

Effective measurement of the RI in liquids is important in biological and chemical research. In the past decade, many different methods have been proposed to perform RI sensing in liquids, such as fiber optic refractometers based on Fabri-Perot interferometers [20], surface plasmon resonance (SPR) refractometers [21, 22], Fano resonance refractometers [23], and photonic crystal (PC) sensors [24–26]. Among them, the PC sensors have attracted a lot of attention because of its rapid colorimetric detection abilities. [24, 25]. The principle of one of the most widely used PC sensors is based on the color transition caused by the change of effective RI on the PC surface. When the solvent molecules are filled into the voids of the PC sensor, the effective RI of the sensor will change, which will be accompanied by a color transition due to the stopband shift in the optical reflection spectrum of the PC sensor. As different solvents cause various stopband shifts, so they can be identified by our eyes through colorimetric detection [26].

In this work, we have fabricated randomly-arranged and hexagonally-close-packed (hcp) dielectric microsphere monolayers made of BTG, PS and silica materials on the top surface of one-dimensional and two-dimensional PC substrates through self-assembly methods, in order to achieve large-area colorimetric and optical RI sensing capabilities in various liquids. The proposed sensor has a rapid colorimetric response to the RI change in liquids. In addition, colorless liquid films with different RI can be easily distinguished according to the structural color of the sensor. By optically measuring the magnification factors of dielectric microspheres, the local RI of liquids in a small area of tens of square microns can be given.

## 2  Materials and methods

### 2.1 Self-assembly of randomly-arranged BTG microsphere monolayers

BTG microspheres (n ~ 1.9 - 2.2) with a mean diameter of 35 $\mu$m were purchased from Microspheres-Nanospheres as powders. They were self-assembled as randomly arranged particles by drop-casting a small amount of BTG powders onto the substrate.

## 2.2   Self-assembly of hcp PS microsphere monolayers

PS microspheres (n ~ 1.59) with 4.9 $\mu$m mean diameters were provided by Nanorainbow, Nanjing as 10 wt% monodispersed suspensions. They were further diluted with ethanol (≥ 99.8%, VWR Chemicals) at a volumetric ratio of 1:4 for colloidal suspension: ethanol. Monolayers of PS microspheres were assembled onto water-air interface with the reported method [27]. First, PS microspheres were applied to water-air interface by drop casting 50 $\mu$L diluted suspension on a tilted glass slide that was partially immersed in deionized water. The PS colloidal suspension immediately spread outwards over water surface because of the surface tension gradient between ethanol-containing colloidal suspension and deionized water. The floating particles in contact began to crystallize and form hcp 2D particle monolayers on the water-air interface. Transferring the floating PS beads from water-air interface to DVD discs was conducted by first immersing the disc in water, then positioning the disc right under the desired region of particle monolayers, and lifting it up to pick the particle monolayers. Finally, the transferred PS microsphere monolayers were dried at room temperature.

## 2.3   Self-assembly of hcp silica microsphere monolayers

A gravity-assisted self-assembly method was used to fabricate hcp silica particle arrays [28, 29]. Silica particles (n ~ 1.45) with a diameter of 0.9 and 4.9 $\mu$m were purchased from Bangs Laboratories, Inc. as 10 wt% colloidal suspensions. They were diluted to 0.67 wt% with deionized water before use. A wedge-shaped capillary cell with three sides open was built by sandwiching two rectangular disc pieces (DVD-R, Philips) on top of a glass substrate at a small angle of ~ 2°. Colloidal suspensions was then injected into the cell and entrapped there due to capillary forces. During experiments, the capillary cell was placed on a bench tilted at 2° forward and 1° to the side, and kept in an environment of 5°C temperature and 60% humidity. A film of silica particle monolayers was formed on the lower side of the capillary cell when the liquid in the suspension dried out.

## 2.4   Microsphere-based RI sensing experiments

The experimental setup for RI sensing is shown in Figure 1 (a). Commercial DVD discs and hcp 900 nm silica particle arrays were used as substrates in this study. The DVD discs have a strip pattern on the top surface, and the separation between the two close tracks is 0.74 $\mu$m [Fig.1 (b)]. The silica particle arrays have a hcp structure [Fig.1 (c)], and they were deposited with 10 nm gold film by thermal evaporation before experiments to improve the image contrast. Various liquids, i.e., water (n=1.33), ethanol (n=1.36), acetone (n=1.36), ethylene glycol (n=1.43) and microscope immersion oil (MIO, n=1.52), were drop casted onto the device with a glass dropper. The pattern of the substrate was collected by self-assembled micrometer-sized microspheres and further magnified through a microscope objective (20 ×, 0.4 NA for 35 $\mu$m BTG microspheres; 50 ×, 0.75 NA for 4.9 $\mu$m PS and silica microspheres; EPIPLAN, Carl Zeiss) in an upright optical microscope system (Axio AX10, Carl Zeiss) illuminated with a halogen lamp (HAL 100, Carl Zeiss) [Fig.1 (d), (e)]. The magnification factor was calculated based on the intensity profiles of the obtained optical microscopic pictures (Figure S1) taken at an axial position of 23 $\mu$m for 35 $\mu$m BTG microspheres and 4 $\mu$m for 4.9 $\mu$m PS and silica microspheres. The Z position is defined as the distance between the top surface of the substrate and the image plane, which was controlled by a homemade high-precision Z-Axis stage.

# 3 Results and Discussion

## 3.1 RI sensing with randomly-arranged BTG microspheres

First, we started by investigating randomly-arranged BTG microspheres. Deionized water and MIO were used as immersion medium and drop-casted onto the PC substrate to submerge the RI sensor. As shown in Figure 2 (a), the strip pattern of the DVD disc was clearly visible when observed through a 35 $\mu$m-diameter BTG microsphere fully immersed in water. When using hcp 900 nm silica particle arrays as PC substrate, an enlarged image of the hcp structures can be observed through the BTG microsphere at an image plane of 23 $\mu$m below the substrate [Fig.2 (b)]. However, we observed some defects in the particle arrays in the image. These defects were caused by the misalignment and the vacancy of colloidal particles during the self-assembly process and cannot be easily eliminated or avoided. Considering that the DVD discs have a better ordering in nanostructure, they were chosen as PC substrates in the RI sensing experiments with PS and silica microsphere arrays. The magnification of the substrate patterns was found to decrease when using MIO as immersion liquid instead of water [Fig.2 (c), (d)]. The colorimetric detection was performed by comparing water and MIO films in the same field of view (FOV) of the optical microscope. As shown in Figure 2 (e), (f), there is a significant difference in the structural color of the PC substrate between the areas immersed in MIO and water. The PC substrate in MIO mostly shows a purple color while it has a light brown color in water [Fig.2 (e)]. As both MIO and water are colorless liquid, the colorimetric difference should be mainly attributed to the stopband shift in the reflection spectrum of the PC substrate. The hcp particle arrays were also found to have a colorimetric response to the RI difference as we observed that the PC substrate in MIO has a darker color than that in water. In addition to the colorimetric difference, the two adjacent BTG microspheres in different liquids clearly showed different magnification factors [Fig.2 (e), (f)].

To investigate this phenomenon in the near-field regime, we performed simulations with Finite-Difference Time-Domain (FDTD) method for 35 $\mu$m-diameter BTG microspheres (n = 1.90). The incident plane waves have 540 nm wavelength. The background medium is assumed to be water (n = 1.33) in Figure 3 (a) and MIO (n = 1.52) in Figure 3 (b). The spatial distribution of light intensity is shown in Figure 3 (a), (b). We found that the FL of the BTG microspheres, which refers to the distance between the PNJ and the center of the microsphere, is longer in MIO than in water.

The correlation between the magnification and the FL of a BTG microsphere is graphically shown in Figure 3 (c). There is a point ($A_1$ or $A_2$) where the parallel beams (blue and orange solid lines) and the center rays (black solid lines) appear to meet after refraction by the lens. This is where the virtual image of the object is formed. For the optical enlargement of an object, a shorter FL ($f_1$ <$f_2$) means a larger virtual image ($h_1$ >$h_2$) and a higher magnification ($h_1/h$>$h_2/h$).

The influence of the RI of immersion liquid on the FL of a 35 $\mu$m-diameter BTG microsphere was studied with FDTD method. As shown in Figure 4 (a), the FL of a 35 $\mu$m-diameter BTG microsphere increases when the background RI increases from 1.0 to 1.6. Therefore, when a liquid with a higher RI is used, it is expected that the magnification of the BTG microsphere will be smaller. Under a constant background RI, it was found that the BTG microsphere with a larger diameter has a longer FL. The correlation between the FL and the microsphere diameter are shown in Figure 4 (b), in which the FL increases linearly with the diameter of the BTG microsphere. Experimental results of the magnification factor *versus* the RI of immersion medium are plotted in Figure 4 (c). Various liquids, i.e., water (n = 1.33), ethanol (n = 1.36), ethylene glycol (n = 1.43) and MIO (n = 1.52) were used to submerge the sensor. We found that the magnification factor decreases from 3.15 × to 2.71 × when the RI of liquid increases from 1.33 to 1.52. Then, experiments were carried out with BTG microspheres of different diameters and water as the immersion medium, in order to study the correlation between the magnification factor and the diameter of the BTG microspheres. As shown in Figure 4 (d), the magnification factor was found to decrease from 4.1 × to 3.1 × as the diameter of BTG microspheres increased from 23 to 38 $\mu$m.

The sensor we developed has an immediate response to RI fluctuations and can be used compatibly with a commercial optical microscope to show RI changes in real time. As shown in Visualization 1 in the supplementary material, there was an obvious change in the strip patterns of the DVD disc when gradually adding ethanol into water, and the structural color of the substrate also turned from dark purple to light brown.

## 3.2  RI sensing with hcp PS and silica microsphere monolayers

In order to perform RI sensing over a large area, centimeter-sized monolayers of hcp 4.9 $\mu$m-diameter PS microsphere arrays were fabricated on a DVD disc by the Langmuir-Blodgett method [27]. The simultaneous existence of regular strip patterns and PS microsphere arrays enables the device capable to perform colorimetric detection and full FOV optical detection of RI differences in liquid. As shown in Figure 5 (a), when using MIO and water to submerge the sensor, the films of the two liquids can be easily distinguished by the color of the sensor. In addition, there is also a difference in the magnification of the strip patterns between MIO and water films [Fig.5 (b)]. As pointed by the arrows in Figure 5 (c)-(e), the boundary between the MIO and water films changed with time. This change cannot be observed due to the colorless nature of the two liquids if using a conventional glass slide as the substrate. However, the evolution of the MIO-water boundary leads to real-time color changes of the sensor, which can be seen by human eyes through an optical microscope.

Silica materials can stand high temperatures and are resistant to organic solvent. In order to study the versatility of this method, monolayers of 4.9 $\mu$m-diameter silica particle arrays were assembled on a DVD disc via the gravity-assisted self-assembly method [28, 29]. Deionized water heated to its boiling point was drop-casted onto the sensor and then gradually cooled down to room temperature. The cooling process was observed with a 50 × microscope objective. As shown in Figure 6 (a), (b), we observed a decrease in the magnification factors of the strip patterns during the cooling process. It is well known that the RI of liquid is higher at low temperatures. Accordingly, this change in magnification should be attributed to the increase in the background RI when the temperature of deionized water decreased. In addition to the change in the magnification factor, the sensor also showed a slight color difference in the area indicated by the white arrows in Figure 6 (a), (b).

## 4   Conclusions

In conclusion, a cost-effective RI sensor comprised of PC substrates and dielectric spherical microlenses has been developed for large-area colorimetric and optical RI sensing applications. The sensor relies on the structural color of the substrate for colorimetric detection, and the sensitivity of the position of PNJ to the RI of the immersion liquid is used for optical detection. Experiments and numerical simulations have been carried out to study the influence of the background RI on the properties of PNJ. We have successfully assembled dielectric microspheres of BTG, PS and silica materials on 1D and 2D PC substrates to perform RI sensing in various liquids. The developed sensor is compatible with commercial optical microscopes and could be used in the applications where various liquids are required, such as microfluidics and biochemical sensing.

## References

1.  Luk'Yanchuk, B. S., Raniagua-Domınguez, R., Minin, I., Minin, O. & Wang, Z. Refractive index less than two: photonic nanojets yesterday, today and tomorrow. *Opt. Mater. Express* **7,** 1821–1847 (2017).

2.  Zhu, J. & Goddard, L. L. All-dielectric concentration of electromagnetic fields at the nanoscale: the role of photonic nanojets. *Nanoscale Adv.* **1,** 4615–4643 (2019).

3.  Lee, J. Y. *et al.* Near-field focusing and magnification through self-assembled nanoscale spherical lenses. *Nature* **460,** 498–501 (2009).

4.  Vlad, A., Huynen, I. & Melinte, S. Wavelength-scale lens microscopy via thermal reshaping of colloidal particles. *Nanotechnology* **23,** 285708 (2012).

5.  Wang, Z. *et al.* Optical virtual imaging at 50 nm lateral resolution with a white-light nanoscope. *Nat. Commun.* **2,** 218 (2011).

6.  Ye, R. *et al.* Experimental far-field imaging properties of a ~5-$\mu$m diameter spherical lens. *Opt. Lett.* **38,** 1829–1831 (2013).

7.  Ye, R. *et al.* Experimental imaging properties of immersion microscale spherical lenses. *Sci. Rep.* **4,** 3769 (2014).

8.  Darafsheh, A., Walsh, G. F., Negro, L. D. & Astratov, V. N. Optical super-resolution by high-index liquid-immersed microspheres. *Appl. Phys. Lett.* **101,** 141128 (2012).

9.  Mahariq, I. *et al.* Photonic nanojets and whispering gallery modes in smooth and corrugated micro-cylinders under point-source illumination. *Photonics* **7,** 50 (2020).

10. Yang, S. *et al.* Influence of the photonic nanojet of microspheres on microsphere imaging. *Opt. Express* **25,** 27551–27558 (2017).

11. Darafsheh, A. Influence of the background medium on imaging performance of microsphere-assisted super-resolution microscopy. *Opt. Lett.* **42,** 735–738 (2017).

12. Liu, X., Hu, S. & Tang, Y. Coated high-refractive-index barium titanate glass microspheres for optically trapped microsphere super-resolution microscopy: A simulation study. *Photonics* **7,** 84 (2020).

13. Ding, H., Dai, L. & Yan, C. Properties of the 3D photonic nanojet based on the refractive index of surroundings. *Chin. Opt. Lett.* **8,** 706–708 (2010).

14. Lee, S. *et al.* Immersed transparent microsphere magnifying sub-diffraction-limited objects. *Appl. Opt.* **52,** 7265–7270 (2013).

15. Wang, F. *et al.* Scanning superlens microscopy for non-invasive large field-of-view visible light nanoscale imaging. *Nat. Commun.* **7,** 13748 (2016).

16. Luo, H. *et al.* Enhanced high-quality super-resolution imaging in air using microsphere lens groups. *Opt. Lett.* **45,** 2981–2984 (2020).

17. Yan, B. *et al.* Superlensing plano-convex-microsphere (PCM) lens for direct laser nano-marking and beyond. *Opt. Lett.* **45,** 1168–1171 (2020).

18. Pacheco-Peña, V., Minin, I. V., Minin, O. V. & Beruete, M. Increasing surface plasmons propagation via photonic nanojets with periodically spaced 3D dielectric cuboids. *Photonics* **3,** 10 (2016).

19. Zhang, T. *et al.* Microsphere-based super-resolution imaging for visualized nanomanipulation. *ACS Appl. Mater. Interfaces* **12,** 48093–48100 (2020).

20. Xu, B., Yang, Y., Jia, Z. & Wang, D. N. Hybrid Fabry-Perot interferometer for simultaneous liquid refractive index and temperature measurement. *Opt. Express* **25,** 14483–14493 (2017).

21. Nasirifar, R., Danaie, M. & Dideban, A. Dual channel optical fiber refractive index sensor based on surface plasmon resonance. *Optik* **186,** 194–204 (2019).

22. Shui, X., Gu, Q., Jiang, X. & Si, G. Surface plasmon resonance sensor based on polymer liquid-core fiber for refractive index detection. *Photonics* **7,** 123 (2020).

23. Zhang, Y. *et al.* High-quality-factor multiple Fano resonances for refractive index sensing. *Opt. Lett.* **43,** 1842–1845 (2018).

24. Endo, T., Yanagida, Y. & Hatsuzawa, T. Colorimetric detection of volatile organic compounds using a colloidal crystal-based chemical sensor for environmental applications. *Sensor Actuat. B: Chem.* **125,** 589–595 (2007).

25. Li, H., Wang, J., Yang, L. & Song, Y. Superoleophilic and superhydrophobic inverse opals for oil sensors. *Adv. Func. Mater.* **18,** 3258–3264 (2008).

26. Hou, J., Li, M. & Song, Y. Recent advances in colloidal photonic crystal sensors: Materials, structures and analysis methods. *Nano Today* **22,** 132–144 (2018).

27. Retsch, M. *et al.* Fabrication of large-area, transferable colloidal monolayers utilizing self-assembly at the air/water interface. *Macromol. Chem. Phys.* **210,** 230–241 (2009).

28. Ye, R., Ye, Y.-H., Zhou, Z. & Xu, H. Gravity-assisted convective assembly of centimeter-sized uniform two-dimensional colloidal crystals. *Langmuir* **29,** 1796–1801 (2013).

29. Fang, C. *et al.* Fabrication of two-dimensional silica colloidal crystals via a gravity-assisted confined self-assembly method. *Colloids Interface Sci. Commun.* **37,** 100286 (2020).

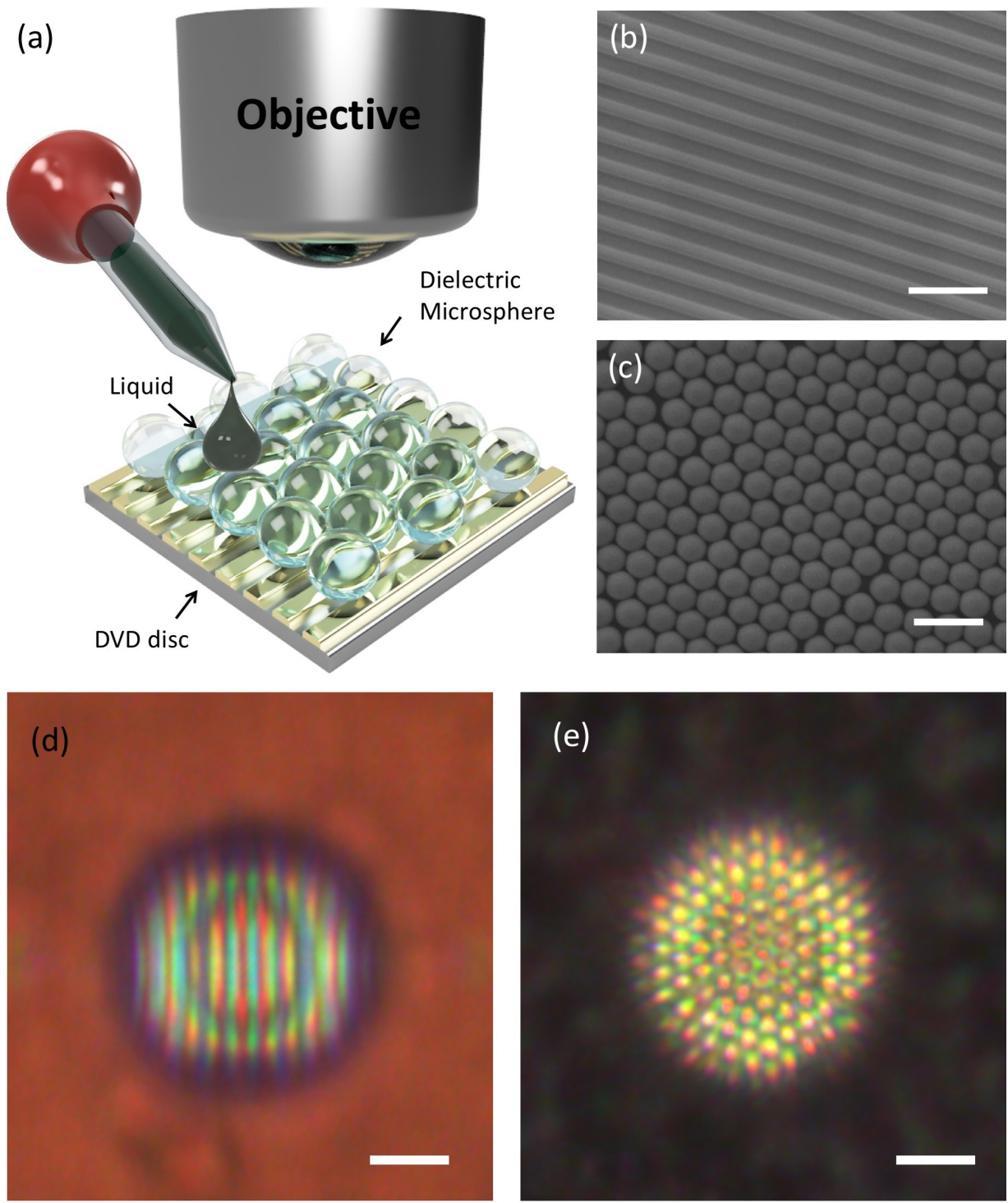

Figure 1: (a) Schematic illustration of the RI sensing setup. (b-c) Scanning electron microscopy (SEM) images of (b) a DVD disc and (c) hcp 900 nm particle arrays. (de) Optical microscopic images of (d) the DVD disc and (e) the particle arrays observed through ethanol-immersed BTG microspheres. The scale bar is 2 $\mu$m in (b), (c) and 10 $\mu$m in (d), (e).

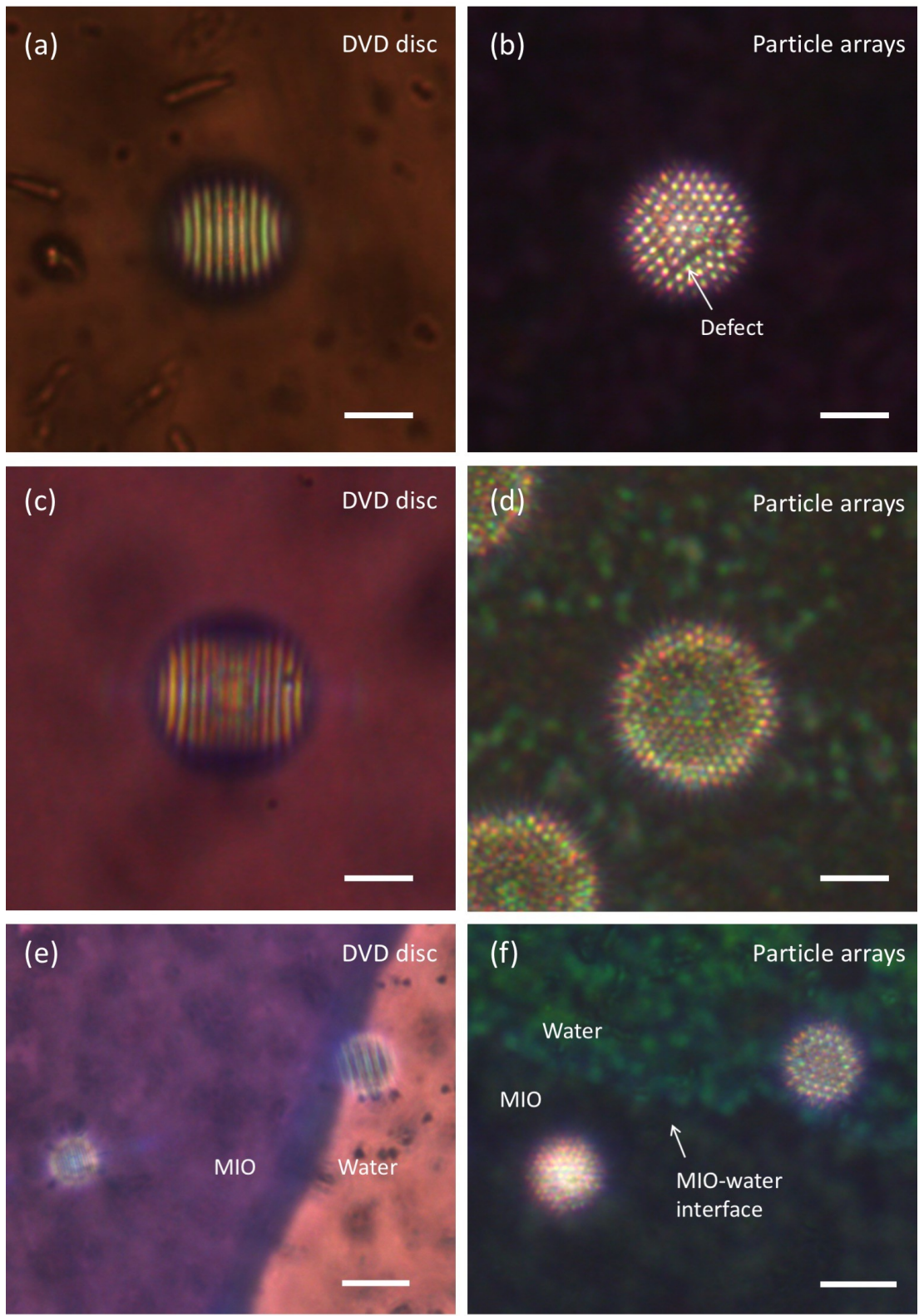

Figure 2: Optical microscopic images of DVD discs and hcp 900 nm particle arrays, in which the patterns of the substrate can be observed through liquid-immersed BTG microspheres. (a-d) The immersion medium is water in (a), (b) and MIO in (c), (d). (e-f) Optical microscopic pictures of the boundary between MIO and water on (e) a DVD disc and (f) particle arrays. The scale bar is 15 $\mu$m in (a)-(d), 40 $\mu$m in (e) and 30 $\mu$m in (f).

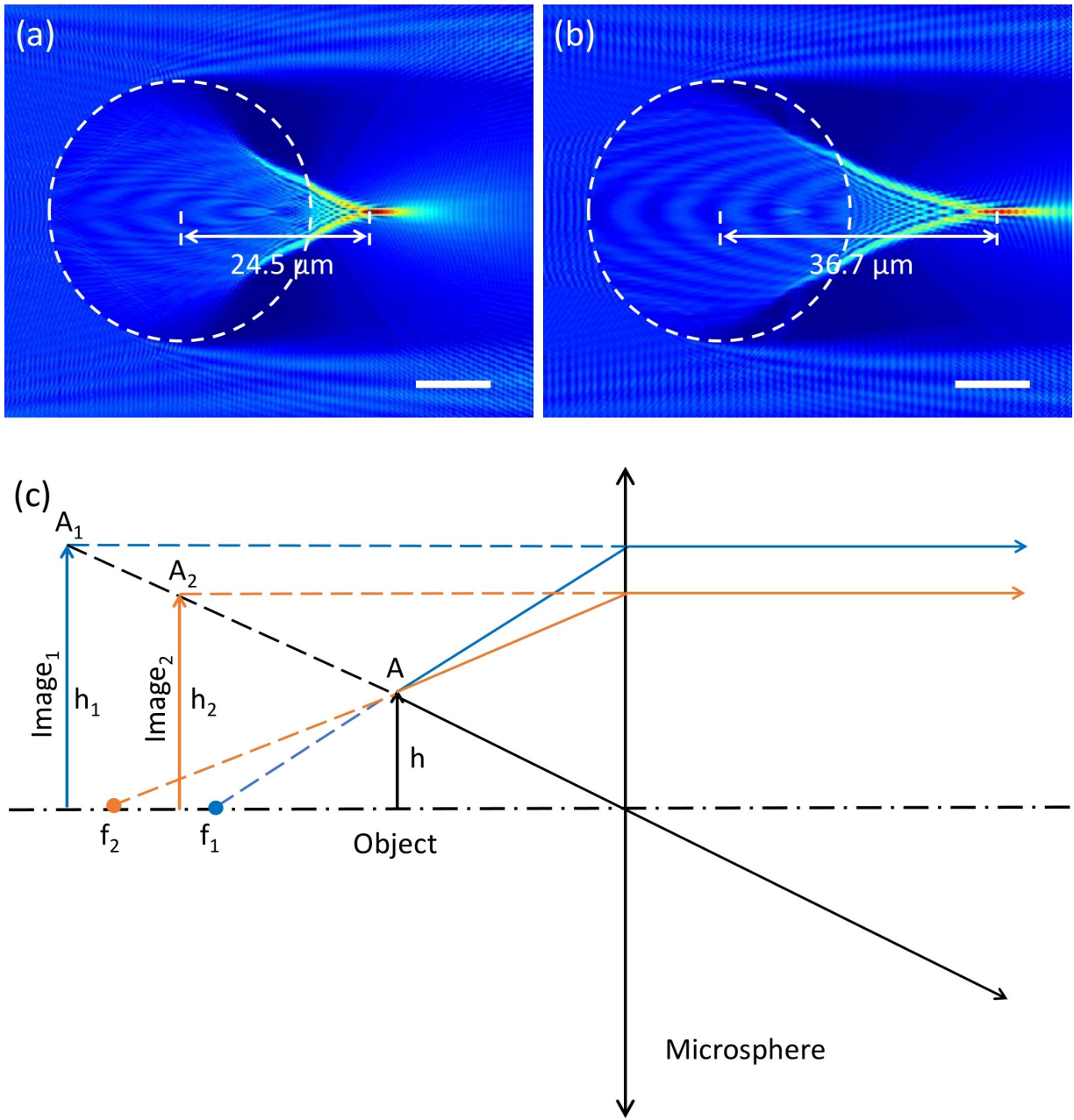

Figure 3: (a-b) Simulation results of the spatial distribution of light intensity adjacent to a 35 $\mu$m-diameter BTG microsphere. The background RI is 1.33 in (a) and 1.52 in (b) MIO. (c) Schematic illustration to show the correlation between PNJ position and magnification factors. The scale bars are 10 $\mu$m in (a), (b).

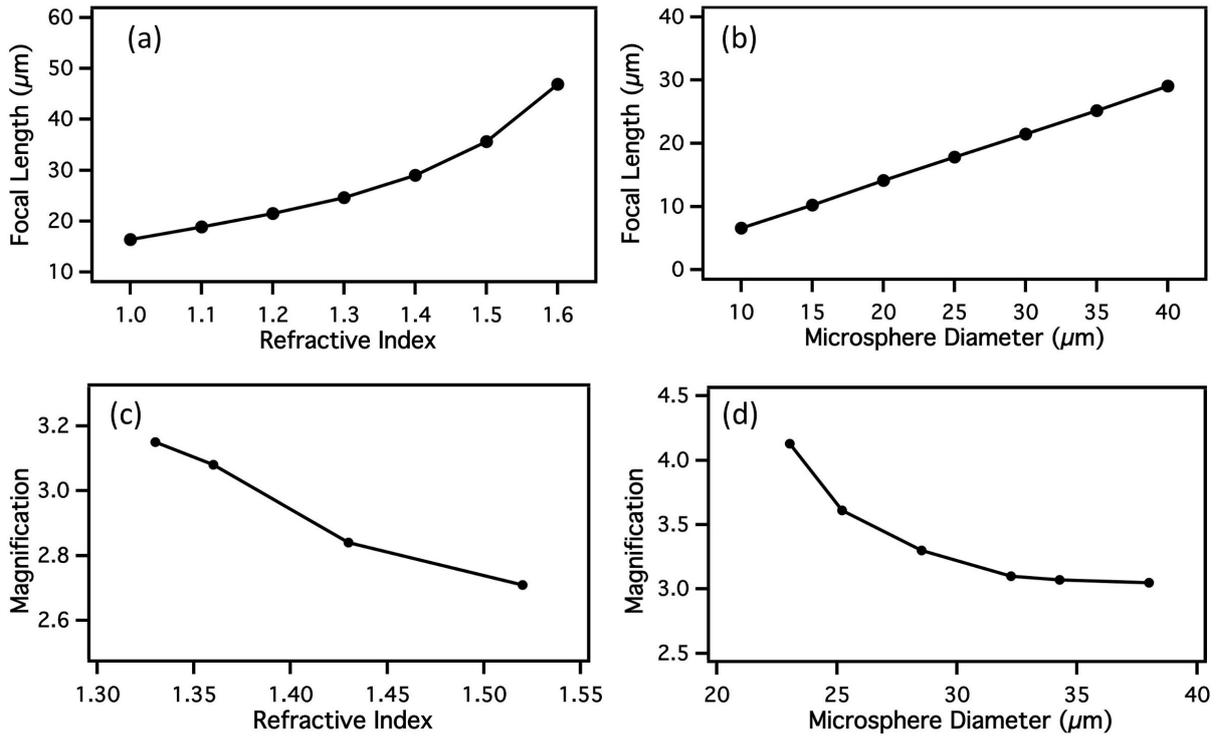

Figure 4: (a) The FL of a 35 $\mu$m-diameter BTG microsphere as a function of the background RI; (b) The FL of water-immersed BTG microspheres as a function of particle diameters; (c) The magnification factors of a 35 $\mu$m-diameter BTG microsphere immersed in liquids with various RI; (d) The magnification factors of water-immersed BTG microspheres with various diameters.

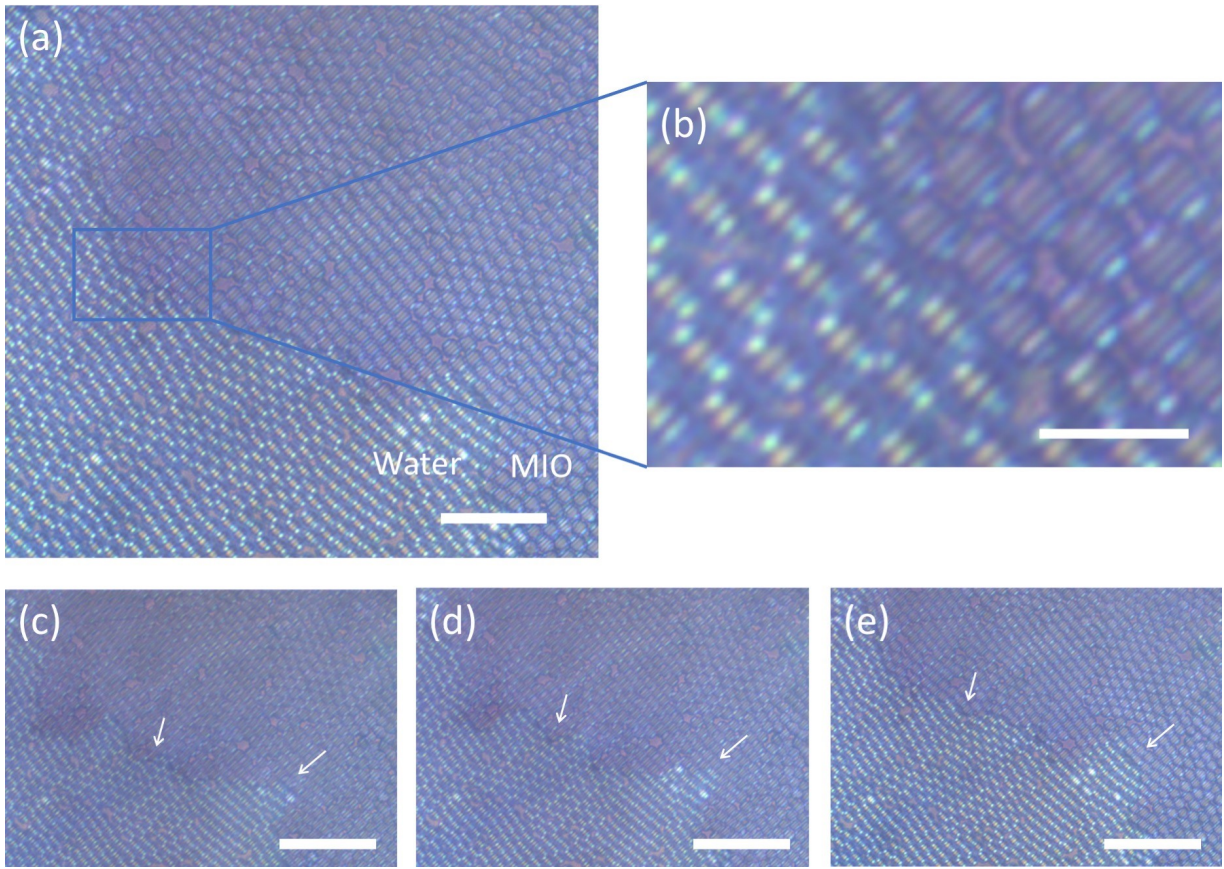

Figure 5: (a) The optical microscopic image of a RI sensor based on hcp 4.9 $\mu$m-diameter PS microsphere monolayers assembled on a DVD disc; (b) Zooming into the blue square; (c-e) The boundary between MIO and water films changes with time. The scale bars are 30 $\mu$m in (a), 10 $\mu$m in (b) and 40 $\mu$m in (c-e).

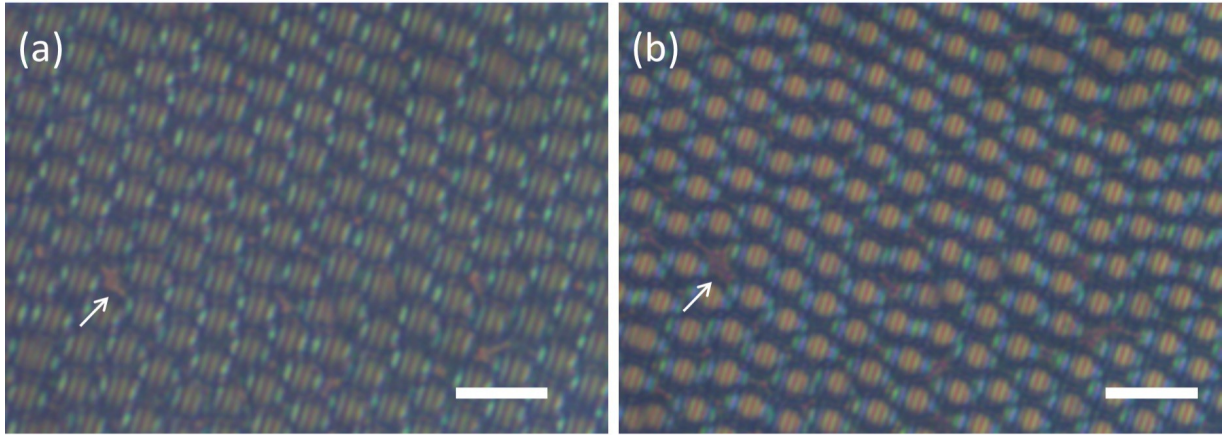

Figure 6: (a) The optical microscopic image of a RI sensor based on hcp 4.9 $\mu$m-diameter silica microsphere monolayers immersed in nearly-boiling deionized water; (b) The optical microscopic image of the same region after the water cooling to room temperature. The scale bars are 10 $\mu$m in (a), (b).